# Acoustic Multifunctional Logic Gates and Amplifier based on Passive Parity-Time Symmetry

Jun Lan,[1] Liwei Wang,[1] Xiaowei Zhang,[1] Minghui Lu,[2] and Xiaozhou Liu[1,*]

[1]Key Laboratory of Modern Acoustics, Institute of Acoustics and School of Physics, Collaborative Innovation Center of Advanced Microstructures, Nanjing University, Nanjing, 210093, P. R. China

[2]National Laboratory of Solid State Microstructures and Department of Materials Science and Engineering, Nanjing University, Nanjing, 210093, P. R. China

Acoustic analogue computation and signal processing are of great significance, however, it's challenging to realize the acoustic computing devices because of their limitations of single function and complex structure. In this paper, an acoustic multifunctional device, which can gate or amplify acoustic waves without resorting to altering the frequency and structure using a passive acoustic parity-time (*PT*)-symmetric metamaterial, is realized theoretically and experimentally. The metamaterial is constructed by five lossless-loss periodically distributed media which are modulated to achieve the passive *PT* symmetry. At the coherent perfect absorber (CPA)-emitter point in the broken *PT*-symmetric phase, the logic gates (AND, OR, XOR and NOT) and small signal amplifier are realized in a single system by adjusting the phase and amplitude differences between two incoming beams, respectively. This work provides a new route for the connection between the *PT* symmetry and the acoustic metamaterial, which has great potential applications in acoustic modulation and acoustic multifunctional device.

[*]Corresponding author: xzliu@nju.edu.cn

## I. INTRODUCTION

In optics, the realization of lasing and anti-lasing in a single system has opened up possibilities for optical multifunctional devices [1,2]. Computational metastructure based on electromagnetic signal has provided more efficient designs for analogy computation [3]. Similar device introduced into acoustics are of vital importance in the development of the efficient information processing and reducing the integration complexity. Acoustic computing devices such as acoustic switches [4,5], filtering [6] and logic gates [7,8] are central elements of acoustic computation and communication, and have attracted extensive attention. However, the realization of multiple functions coexisting without resorting to altering the frequency and structure in these acoustic computing devices is a significant challenge. Many acoustic computing devices have been proposed relying on phononic metamaterials (PMs), which can control the propagation states of acoustic waves by using the band gaps and distinct frequency characteristics [9,10]. So far, acoustic computing devices based on PMs have be realized by means of the self-collimation effect [7], control waves [11], and rotating the rods of PMs [12]. However, most of them suffer simple functionality or complex design. Therefore, it is necessary to seek for a simple and effective approach to design acoustic multifunctional devices that can tackle above issues.

Meanwhile, parity-time ($PT$) symmetry is initially proposed as a concept in the area of quantum mechanics [13,14]. It has shown that the special class of Hamiltonians that commute with $PT$ operator can exhibit entirely real energy eigenvalues even though they are non-Hermitian. In recent years, $PT$ symmetry is extended to classical wave system and its counterparts in the fields of optics, electronics and acoustics have been also found [15-20], which has broaden the approach to wave manipulation. Exceptional point (EP) is a singular point in the breaking *PT*-symmetric phase, where a spontaneous *PT* symmetry breaking can occur and one-way zero reflection is arising [21-23], A flurry of recent studies have demonstrated phenomena associated with the unique property of the EP [24-26]. Novel phenomena, such as unidirectional invisibility [16,27], reflectionlessness [28], coherent perfect absorption [29,30] and single-mode lasing [31-33], have been successively explored. Recently, some researches have emphasized the existence of the novel singular point in the broken *PT*-symmetric phase in optics, which refers to coherent perfect absorber (CPA)-laser point [1,2,34-36]. The CPA-laser point occurs when a pole and a zero of the scattering matrix coexist at a specific frequency. The *PT*-symmetric system can function simultaneously as a CPA and a laser at threshold, which offers a new route for light modulation with high contrast approaching the ultimate limit. In acoustics, to realize *PT* symmetry in acoustic metamaterials, one must overcome the challenge associated with the absence of acoustic gain medium in nature [37]. The passive acoustic *PT*-symmetric metamaterials have recently been proposed to realize the asymmetric wave behaviors at the EPs [38-41]. Existing works on acoustic *PT* symmetry mainly paid attention to the scattering properties around the EPs. If the optical concept of CPA-laser point could introduce into the acoustic *PT*-symmetric system, a multifunctional acoustic device with capabilities of an amplifier, an absorber and a modulator is expected to realize.

In this paper, the concept of CPA-emitter point is proposed in passive acoustic *PT*-

symmetric metamaterial, which is the acoustic counterpart of CPA-laser point. The alternating lossless-loss modulation of the passive *PT*-symmetric potential is introduced by the rectangular waveguide with periodically distributed leakage structures. The lossless and loss regions of the passive metamaterial are carefully engineered to approach highly efficient coherent control with a maximum contrast between the absorption and emission states. At the CPA-emitter point where a pole and a zero of the scattering matrix coexist, we theoretically and experimentally realize both logic gates and amplifier by controlling the initial phase and amplitude differences of the two incident waves, respectively.

## II. THEORETICAL DESIGN OF THE PASSIVE ACOUSTIC *PT*-SYMMETRIC SYSTEM

For our design, as depicted in Fig. 1(a), the passive *PT*-symmetric modulation is evolved from the exact *PT*-symmetric modulation. As shown in the top panel of Fig. 1(a), the classic CPA-laser *PT*-symmetric system for light has simple harmonic modulation of the refractive index $n(x) = n_0 + in_i \sin(\pi x/L)$, where $L$ is the Bragg period corresponding to half of the wavelength, $n_0$ is the background refractive index and $n_i$ is the modulation index of the imaginary part. This harmonic modulation is further simplified by a simple square-wave modulation, as shown in the middle panel of Fig. 1(a). By introducing the concept of CPA-laser *PT* symmetry into passive acoustic *PT* symmetry, the square-wave modulation is truncated by only considering the imaginary part modulation with negative value, as shown in the bottom panel of Fig. 1(a). The revised passive acoustic *PT*-symmetric modulation of the refractive index is

expressed as

$$\begin{cases} n(x) = n_0 & (m-3.5)L \leq x \leq (m-3)L \\ n(x) = n_0 - n_i i & (m-3)L \leq x \leq (m-2.5)L \end{cases} \quad (1)$$

where $m = 1,2,3,4,5$, $L = \lambda_0/2 = 50$ mm, and $n_i$ is tuned from 0 to positive value for the loss medium. The modulation period is still unchanged through the evolution process. Here, the operating frequency is the Bragg frequency $f_B = 3430$ Hz and the corresponding CPA-emitter point of the passive *PT*-symmetric system is approached at this frequency. The passive *PT*-symmetric system is composed of a ten-layer structure with lossless (*A*) and loss regions (*B*) arranged alternatively and periodically, as shown in Fig. 1(b).

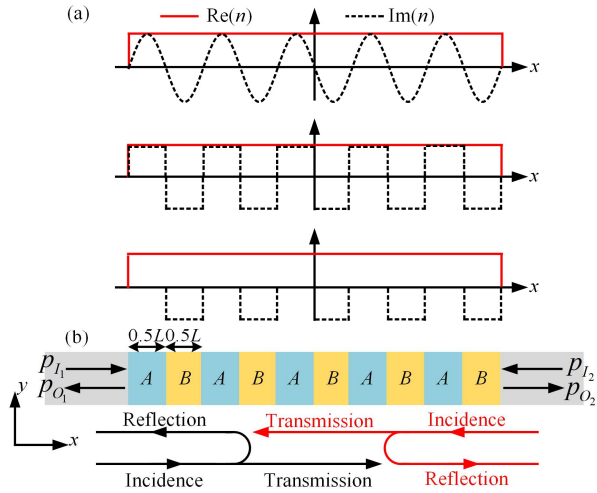

FIG. 1. Design of the passive acoustic *PT*-symmetric system. (a) Evolution of *PT*-symmetric potential. The top and middle panels represent the exact *PT*-symmetric potential with simple harmonic modulation and square-wave modulation, respectively. The bottom panel represents the revised passive *PT*-symmetric potential. (b) Arrangement of the lossless (*A*) and loss (*B*) regions of the passive acoustic *PT*-symmetric system.

## III. OPERATING PRINCIPLE AND STRUCTURE FABRICATION

At the CPA-emitter point, the passive acoustic *PT*-symmetric system can support both coherent perfect absorption and emission states coexistence with maximum contrast, which is similar with CPA-laser point in optics and has a benefit to construct multifunction acoustic device [1]. As shown in Fig. 1(b), the passive acoustic *PT*-symmetric system is a two-port system. The transfer matrix method is used to derive the acoustic scattering matrix *S* describing the relation between the input and output acoustic waves, i.e., $\begin{pmatrix} O_1 \\ O_2 \end{pmatrix} = \begin{pmatrix} t & r_L \\ r_R & t \end{pmatrix} \begin{pmatrix} I_2 \\ I_1 \end{pmatrix} = S \begin{pmatrix} I_2 \\ I_1 \end{pmatrix}$, where $I_1$ and $I_2$ are the amplitudes of the incident waves from the left and right ports, respectively. $O_1$ and $O_2$ are the amplitudes of the output waves at the left and right ports, which are the superpositions of the transmitted and reflected waves. $r_{L(R)}$ and *t* are the left-(right-) reflection and transmission coefficients, respectively. *t* is the transmission coefficient, which is identical for both left and right incident waves due to reciprocity. At the singular CPA-laser point of the exact *PT*-symmetric system, the absolute value of one eigenvalue of scattering matrix goes to pole, while the other goes to zero. In the passive acoustic *PT*-symmetric system, the CPA-emitter point should have the identical property. Here, the two associated eigenvalues of the scattering matrix *S* are derived as $\lambda_{1,2} = t \pm \sqrt{r_L r_R}$. We then use the transfer matrix method to estimate the suitable value of the modulation index of the imaginary part ($n_i$) in Eq. (1). At the operating frequency 3430 Hz, the absolute values of two eigenvalues $|\lambda_{1,2}|$ as a function of $n_i$ is shown in Fig. 2(a). It is seen that, the near zero value of the $|\lambda_2|$ is only achieved when $n_i \approx 0.167$, which corresponds to the vicinity of CPA-emitter point. In the case of $n_i = 0.167$, Fig. 2(b) shows the absolute values of two eigenvalues $|\lambda_{1,2}|$ as a function of frequency, where the zero ($\approx 0.02$) and the pole of two eigenvalues in the broken *PT*-symmetric phase ($\approx 0.68$) coexist around the operating frequency 3430 Hz. It is worth mentioning that there is a slight deviation between the actual frequency of the CPA-emitter point (3400

Hz) and the operating frequency (3430 Hz), which is within the acceptable range. Therefore, under the condition $n_i = 0.167$, the proposed passive *PT*-symmetric system could display exotic property of CPA-emitter point. The amplitudes of the transmitted, and reflected waves at the left and right ports as a function of frequency are presented in Fig. 2(c). It shows that the left and right reflection and transmission coefficients are all around 0.34 at the operating frequency. The corresponding phases of the transmitted waves and the left- and right-reflected waves are shown in Fig. 2(d). At the operating frequency, there is $\pi$ phase difference between two reflections, and the phase difference between the transmission and reflections of the two ports is $\pm \pi/2$. Therefore, at the CPA-emitter point, the phase-sensitive approaches the ultimate limit in the broken *PT*-symmetric phase. By adjusting the phase difference between two incident acoustic waves, the logic gates such as AND, OR, XOR and NOT can be achieved. Figures 2(a)-2(d) indicate that the CPA-emitter point creates two mutually opposite states: when the phase difference between the two beams is $\Delta\phi = \phi_1 - \phi_2 = -\pi/2$, acoustic wave becomes localized in the loss regions, and the amplitudes of the output acoustic waves of two ports are approximately zero, corresponding to the output logic state 0. This presents the foundation to realize XOR and NOT gates with high contrast ratio. Here $\phi_1$ and $\phi_2$ are the initial phases of the left and right incident beams, respectively. When the phase difference between two beams is $\Delta\phi = \pi/2$, acoustic wave is interferometrically confined in the lossless regions and the outgoing beams of two ports are resonantly emitted. By appropriately selecting the thresholds, the proposed device can function as AND and OR gates. A small signal amplifier can be realized through defining the left and right ports as the control port and the incident port, respectively.

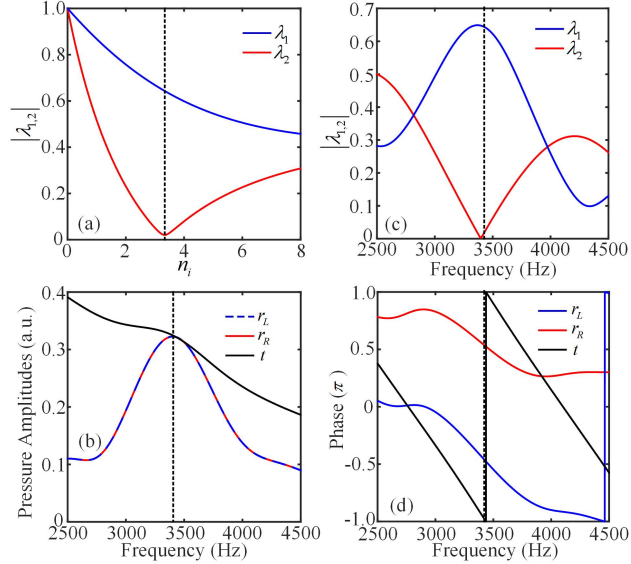

FIG. 2. Characteristics of the acoustic *PT*-symmetric system. (a) and (b) Absolute values of two eigenvalues as functions of the values of $n_i$ and frequency, respectively. (c) Pressure amplitudes of the left-(right-) reflected and transmitted waves of the passive *PT*-symmetric system as a function of frequency, respectively. (d) Corresponding phases of the left-(right-) reflected and transmitted waves as a function of frequency. Vertical dotted line in (a) indicates $n_i = 0.167$, Vertical dotted lines in (b), (c) and (d) indicate the frequency $f = 3430$ Hz.

We have presented that a well-designed passive acoustic *PT*-symmetric system possesses physical features of CPA-emitter point. We further replace the lossless and loss media in Fig. 1(b) with waveguide metamaterial, as shown in Fig. 3(a). According to Eq. (1), the material parameters (mass density and acoustic velocity) for the lossless *A* and loss *B* regions are $\rho_A = 1.21 \text{ kg/m}^3$, $c_A = 343$ m/s, $\rho_B = 1.21 \text{ kg/m}^3$ and $c_B = 333.72 + 55.648i$ m/s, respectively. Thus, *A* modulation can be achieved by introducing an acoustic rectangular waveguide without any decoration. The length, width and thickness of the cross section of the rectangular waveguide are $a = 40$ mm, $b = 40$ mm and $d = 5$ mm, respectively. For the loss medium *B* with refractive index $n_B = n_0 - 0.167i$, a deliberate control of sound loss is required. Generally, the attenuation effect can be obtained by the purely resistive leakage coming from the vent slits [39]. The cross section of *B* modulation in *x-z* plane is shown in Fig. 3(b). In the full-wave simulation, the leakage effect can be included with acoustic impedance. For simplicity, the impedance boundary conditions are directly utilized to describe the upper and lower boundaries of the rectangular waveguides in the simulation, so that the complex structures can be easily replaced by the acoustic impedance homogeneously distributed at the boundaries. Based on the retrieval method of obtaining effective properties of the acoustic metamaterial [42], the simulated real part (black solid line) and imaginary part (black dotted line) of the refractive index of the loss medium *B* as a function of boundary acoustic impedance are displayed in Fig. 3(c). When the boundary acoustic impedance is $410 \text{ Pa} \cdot \text{s/m}$, the refractive index is about $n = 0.95 - 0.161i \approx n_B$, where the real and imaginary parts of the refractive index are close to 1 and 0.167, respectively. Therefore, the passive acoustic *PT*-symmetric metamaterial can be constructed by a rectangular waveguide with leakage structures periodically decorated on the waveguide boundaries.

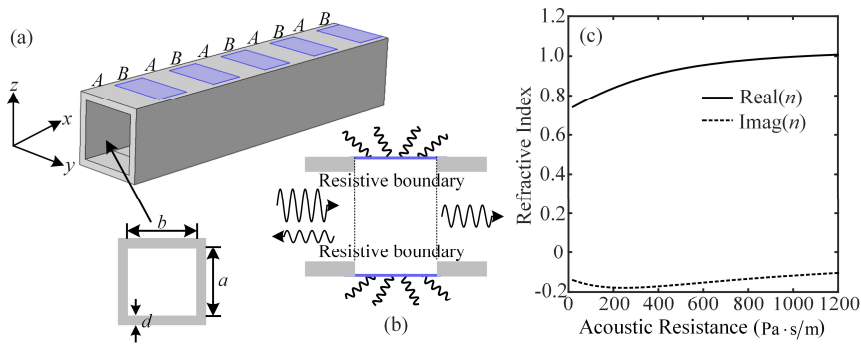

FIG. 3. (a) Schematic of the passive acoustic waveguide metamaterial combining the lossless *(A)* and loss *(B)* media. (b) The cross-sectional picture of the acoustic metamaterial for the loss medium in *x-z* plane. Resistive boundary is equivalent to leakage structure. (c) Simulated real and imaginary parts of the refractive index of the loss medium (*B*) at the operating frequency 3430 Hz.

## IV. SIMULATION AND EXPERIMENTAL RESULTS

We further perform full-wave simulations (COMSOL Multiphysics) to verify the *PT*-symmetric characteristic at the CPA-emitter point of the designed waveguide metamaterial and realize the functions of acoustic logic gates and amplifier. For the case of the left and right inputs ($p_0$, 0) or (0, $p_0$), only one signal beam with amplitude $p_0$ is injected into left or right port of the metamaterial, the simulated output pressure amplitudes of the left and right ports at the operating frequency are shown in Fig. 4(a) or 4(b). The results indicate that the output pressure amplitude of the left port is nearly equal to that of the right port for both cases (around $0.34 p_0$ and $0.37 p_0$), which is consistent with the theoretical results shown in Fig. 2(c). For the case of left and right inputs ($p_0$, $p_0$) with initial phase difference $\Delta\phi = \pi/2$ or $\Delta\phi = -\pi/2$, the corresponding result is illustrated in Fig. 4(c) or 4(d). At the operating frequency 3430 Hz, the less absorption and strong output signal (around $0.7 p_0$) is observed when $\Delta\phi = \pi/2$. In contrast, the coherent perfect absorption is achieved with almost no output signal (around $0.04 p_0$) when $\Delta\phi = -\pi/2$. Hence, with the designed passive acoustic *PT*-symmetric metamaterial, the absorption and emission states can be engineered to be coexist, which could be utilized to realize the logic gates and amplifier. From the basic principle of acoustic wave interference, there should set two thresholds for realizing the logic gates (AND, OR, XOR and NOT). As shown in Figs. 4(a) and 4(c), when acoustic waves with amplitude $p_0$ are applied at the left port and the two ports (left and right ports), the transmission peak at the output are around $0.34 p_0$ and $0.68 p_0$, respectively. The threshold $A_1 = 0.34 p_0$ is set to realize functions of XOR and OR gates and the threshold $A_2 = 0.68 p_0$ is set to realize the functions of AND and NOT gates. First, as shown in Figs. 4(a), 4(b) and 4(c), when a threshold of $A_1$ is applied, the output logic state 1 responses to inputs ($p_0$,0), (0, $p_0$) and ($p_0$, $p_0$), which correspond to input logic states (1,0), (0,1) and (1,1). Hence, the proposed device functions as an OR gate. When a threshold of $A_2$ is applied, the output logic state 1 responses to input logic state (1,1), while the output logic states are 0 for the input logic states (1,0) and (0,1). As a result, the AND gate is realized. Moreover, as shown in Figs. 4(a), 4(b) and 4(d), when a threshold of $A_1$ is applied, the output logic state 1 responses to input logic states (1,0) and (0,1), while the output logic state is 0 for the input logic state (1,1). Therefore, the proposed device functions as a XOR gate. The NOT gate is realized by choosing the left port as a control port with fixed amplitude $p_0$ and the right port as an input port. As shown in Figs. 4(a) and 4(d), with the threshold of $A_1$, when the right input is turned off, the output logic state is 1, while the output logic state is 0 for the right input pressure amplitude $p_0$.

Moreover, by using the same phase difference as the AND logic gate, a small signal amplifier can be realized when the left port is defined as the control port and the right port is defined as the incident port, as shown in Fig. 4(e). Here, the amplification coefficient for the device is characterized by $O_2/I_2 = 0.34(1/(I_2/I_1)+1)$. The theoretical result for the small signal amplifier is shown in Fig. 4(f) with black solid curve. When the input signal $I_2$ is a small signal relative to the control signal $I_1$ (around $I_2/I_1 < 0.5$), the ratio of output signal $O_2$ to input signal $I_2$ is more than one, which corresponds to the working zone of amplifier. We refer to this system as the small signal amplifier device. In an extreme case of $I_2/I_1 = \infty$, $O_2/I_2$ verges to 0.34. The red triangles in Fig. 4(f) are the numerical results of amplifier, which are consistent with the theoretical results. Therefore, the proposed passive acoustic $PT$-symmetric metamaterial can function as AND, OR, XOR and NOT gates or small signal amplifier without resorting to altering the frequency and structure.

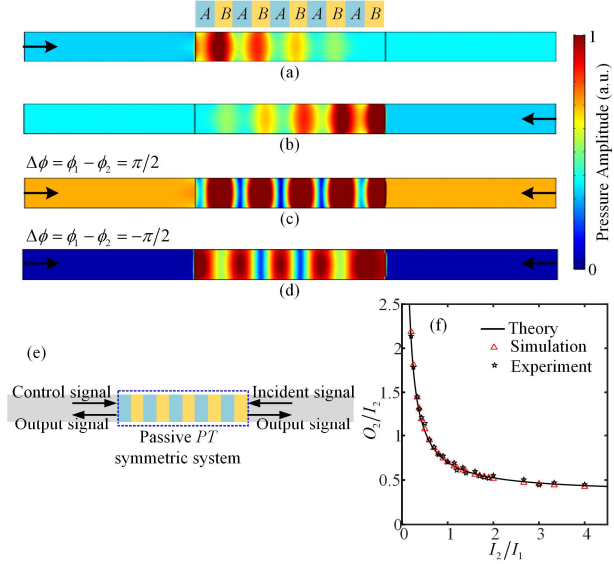

FIG. 4. Behaviors of the acoustic logic gates and amplifier. Pressure amplitude distributions inside and outside the $PT$-symmetric metamaterial under only one incident wave from (a) left port and (b) right port, and under two incident waves with phase differences (c) $\Delta\phi = \pi/2$ and (d) $\Delta\phi = -\pi/2$, respectively. The pressure amplitude distributions outside the metamaterial have subtracted the distribution of incident wave. (e) Schematic of the passive $PT$-symmetric metamaterial for the small signal amplifier. (f) Theoretical, simulated and measured amplification coefficient $O_2/I_2$ versus input contrast ratio $I_2/I_1$.

In the following, the experimental measurements are performed to verify above theoretical and simulated results. The set-up of the experiment is shown in Fig. 5(a), where the propagating medium of sound wave is air and sound-absorbing cottons are placed at both ends of the waveguide to eliminate the reflection. Figure 5(b) is the sample of the passive $PT$-symmetric metamaterial. We choose the mesh fabric, which are woven with monofilament fibers (Saatifil Acoustex HD15), as the leakage structure. The thickness, average pore and open area of this mesh fabric are $55\mu m$, $15\mu m$ and 9%, respectively, and the effective acoustic impedance is about $410 \, \text{Pa} \cdot \text{s/m}$. This experimental system can realize the functions of the acoustic logic gates and the small signal amplifier. Here, two speakers are used to generate the left and right input signals ($p_{I_1}$ and $p_{I_2}$) from function generator. The output signals of the left and right ports ($p_{O_1}$ and $p_{O_2}$) are picked up and amplified by two microphones and signal conditioner, then sent them to PC by the oscilloscope. What needs to be illustrated is that, since the pressure amplitudes of the left and right output signals are almost same at the operating frequency 3430 Hz, the normalized output signal of the left port $p_{O_1}$ is only presented in experiment results of Figs. 6(c)-6(e). The truth table of all functions of the logic gates are presented in Table 1. The combination of two input signals as shown in Figs. 6(a) [I → IV] and 6(b) [I → IV] goes through all the four input states (1,1), (0,1), (1,0) and (0,0) with time 2 ms. In Figs. 6(a) (I) and 6(b) (I), the phase difference $\Delta\varphi$ between two inputs shown by red curves is $\pi/2$ and between two inputs shown by black curves is $-\pi/2$. For instance, for an AND gate, the phase difference $\Delta\varphi$ is set as $\pi/2$, and the two states of the speakers' switch (on/off) represent the binary states (1/0). If the threshold value is set to $A_2$ (black dotted line), the left output signals of the passive *PT*-symmetric metamaterial present four logic states (1, 0, 0, 0) orderly resembling the AND gate. Similarly, when the phase difference remains unchanged ($\Delta\varphi = \pi/2$) and another threshold value $A_1$ (red dotted line) is set, the output signals give four different states (1, 1, 1, 0) which function as OR gate. Another example is that when the phase difference is chosen as $\Delta\varphi = -\pi/2$ and the threshold value is set to $A_1$, the third kind of four output signal states (0, 1, 1, 0) can be achieved, in turn functioning as XOR gate.

In a special case, if the threshold remains as $A_1$, and $p_{I_1}$ is set as the control signal with fixed state 1 (I and III, Figs. 6(a)) and $p_{I_2}$ is set as the input signal (I and III, Figs. 6(b)), the corresponding output signal can work as an NOT gate, as shown in I and III, Figs. 6(e). The experimental results of the small signal amplification capability are shown by the black stars in Fig. 4(f), which are consistent with the theoretical results (black solid curve). The experimental results indicate that the passive *PT*-symmetric metamaterial can successfully act as an acoustic multifunctional device with the capabilities of acoustic logic gates and small signal amplifier.

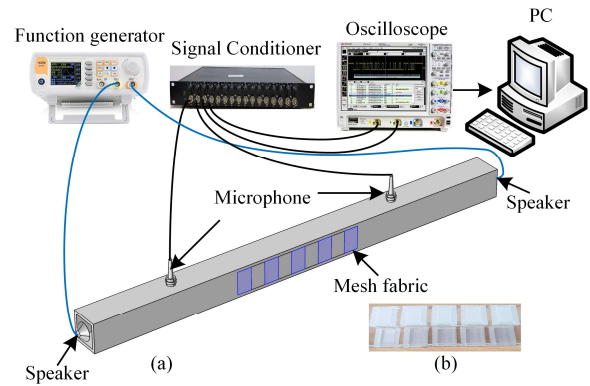

FIG. 5. Experimental arrangement and passive *PT*-symmetric metamaterial prototype. (a) Schematic of the experimental arrangement. (b) The photograph of sample fabricated by a rectangular waveguide with leakage structures periodically decorated on the waveguide boundaries.

Table 1 Truth tables of AND, OR, XOR and NOT gates

| Logic Gates $I_1$ | $I_2$ | AND | OR | XOR | NOT |
|---|---|---|---|---|---|
| 1 | 1 | 1 | 1 | 0 | 0 |
| 0 | 1 | 0 | 1 | 1 |  |
| 1 | 0 | 0 | 1 | 1 | 1 |
| 0 | 0 | 0 | 0 | 0 |  |

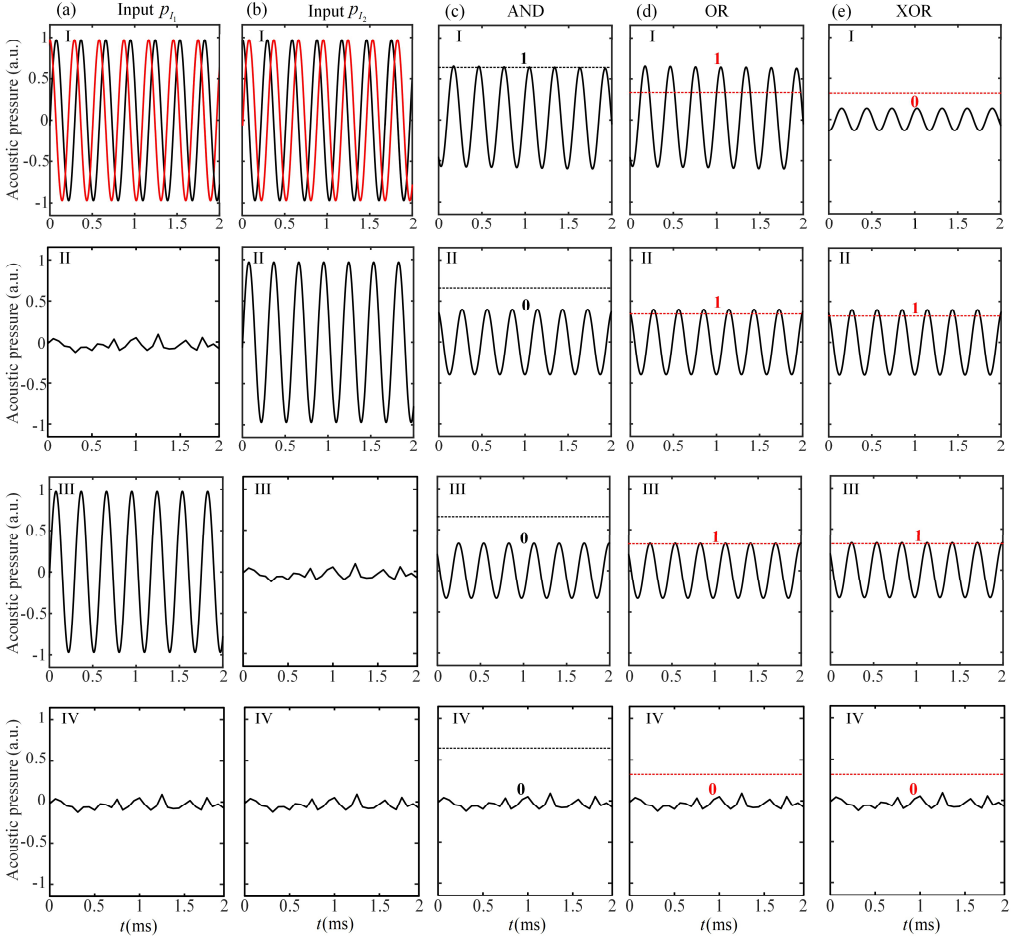

FIG. 6. Measured signal for the logic gates of the passive acoustic $PT$-symmetric metamaterial at the operating frequency 3430 Hz. (a) I-IV and (b) I-IV left and right input signals $p_{I_1}$ and $p_{I_2}$, respectively. (c) I-IV, (d) I-IV, and (e) I-IV left output signal $p_{O_1}$ for AND, OR and XOR gates, respectively. The red and black dotted lines mark $A_1$ and $A_2$, respectively.

## V. CONCLUSION

This work presents a realization of a multifunctional device, controlled by $PT$ symmetry, which offers logic gates or small signal amplifier without resorting to altering the frequency and structure. An efficient and accurate analysis method has been put forward for the design of the passive acoustic $PT$-symmetric metamaterial with exotic scattering properties at the CPA-emitter point. Through balancing the interplay between the lossless and loss media, the logic gates and small signal amplifier are realized by controlling the phase and amplitude differences of the input signals, respectively. In the following work, the integration of acoustic delay line enabling the phase difference will be accurately controlled and may promise highly integrated acoustic multifunctional devices, which will offer a simple and effective approach for reducing the integration complexity and signal manipulation in acoustic communication.

## ACKNOWLEDGMENT

J. L. and L. W. W contributed equally to this work. This work was supported by the National Key R&D Program of China (No. 2017YFA0303702), State Key Program of National Natural Science Foundation of China (No. 11834008), National Natural Science Foundation of China (No. 11774167), State Key Laboratory of Acoustics, Chinese Academy of Science (No. SKLA201809), Key Laboratory of Underwater Acoustic Environment, Chinese Academy of Sciences (No. SSHJ-KFKT-1701) and AQSIQ Technology R&D Program, China (No. 2017QK125).